\documentclass[conference]{IEEEtran}

\usepackage{cite}
\usepackage{amsmath,amssymb,amsfonts}
\usepackage{algorithmic}
\usepackage{graphicx}
\usepackage{textcomp}
\usepackage{xcolor}
\usepackage{paralist}
\usepackage{url}
\usepackage[hidelinks]{hyperref}

\makeatletter
\def\ps@IEEEtitlepagestyle{%
\def\@oddfoot{\parbox{\textwidth}{\footnotesize
Authors' version of a paper accepted for publication in the proceedings of the 1st ITG Workshop on IT Security. 
Cite as: L.~Roepert, M.~Dahlmanns, I.~B.~Fink, J.~Pennekamp, and M.~Henze, ``Assessing the Security of OPC UA Deployments'', in \emph{Proceedings of the 1st ITG Workshop on IT Security (ITSec)}, 2020.\vspace{3em}}
}%
}
\makeatother

\begin{document}

\title{Assessing the Security of OPC UA Deployments}

\author{%
\IEEEauthorblockN{%
Linus Roepert\IEEEauthorrefmark{1},
Markus Dahlmanns\IEEEauthorrefmark{1},
Ina Berenice Fink\IEEEauthorrefmark{1},
Jan Pennekamp\IEEEauthorrefmark{1},
Martin Henze\IEEEauthorrefmark{3}
}
\IEEEauthorblockA{\IEEEauthorrefmark{1}\textit{Communication and Distributed Systems}, \textit{RWTH Aachen University}, Aachen, Germany, \{lastname\}@comsys.rwth-aachen.de}
\IEEEauthorblockA{\IEEEauthorrefmark{3}\textit{Cyber Analysis \& Defense}, \textit{Fraunhofer FKIE}, Wachtberg, Germany, martin.henze@fkie.fraunhofer.de}
}

\maketitle

\begin{abstract}
To address the increasing security demands of industrial deployments, OPC UA is one of the first industrial protocols explicitly designed with security in mind.
However, deploying it securely requires a thorough configuration of a wide range of options.
Thus, assessing the security of OPC UA deployments and their configuration is necessary to ensure secure operation, most importantly confidentiality and integrity of industrial processes.
In this work, we present extensions to the popular Metasploit Framework to ease network-based security assessments of OPC UA deployments.
To this end, we discuss methods to discover OPC UA servers, test their authentication, obtain their configuration, and check for vulnerabilities.
Ultimately, our work enables operators to verify the (security) configuration of their systems and identify potential attack vectors. 
\end{abstract}

\section{Introduction}

Traditionally, industrial protocols, such as Modbus and Profinet, were designed to operate in isolated networks and thus provide little to no security functionality.
However, the increasing interconnection of industrial processes~\cite{pennekamp_towards_2019} as well as serious security threats, e.g., evidenced by the cyberattack on a German steel mill or NotPetya~\cite{hemsley_history_2018}, demand for \emph{secure} industrial protocols.
One prime candidate to fill this demand is OPC Unified Architecture (OPC UA)~\cite{opcua_specification_2017}, a comparatively new protocol enabling standardized and secure communication across all levels from the field up to the cloud. 

OPC UA has been explicitly designed with security in mind as attested by a security analysis performed on behalf of the German Federal Office for Information Security~\cite{bsi_opcua_2017}.
Still, OPC UA is not secure by default, requiring a thorough configuration~\cite{opcua_security_2018}.
OPC UA's inherent complexity and a wide range of deployment models make this setup a laborious and error-prone task.
In fact, several setup instructions for OPC UA servers can result in insecure deployments~\cite{angeli_secure_2018,bsi_opcua_2017}.
Consequently, a need for assessing the security of OPC UA deployments exists, e.g., when integrating them into industrial environments.
This is especially crucial for deployments which are (un-)intentionally connected to the Internet, where several thousand probes are searching for exploitable deployments~\cite{fachkha_probing_2017}.

To aid in performing network-based security assessments of OPC UA deployments, we present approaches to
\begin{inparaenum}[(i)]
\item discover OPC UA servers in a network,
\item test for anonymous, default, or weak login credentials,
\item retrieve information on the server and security configuration as well as access rights, and
\item test for susceptibility to certain known CVEs and other potential vulnerabilities.
\end{inparaenum}
We integrate our tools into the widely used Metasploit Framework and release the source code\footnote{Available as open source at \url{https://github.com/COMSYS/msf-opcua}}.

\section{OPC UA Security}
\label{sec:background}

OPC UA servers represent objects and their relationships as a set of nodes in an address space.
To realize security, OPC UA provides authentication, authorization, as well as integrity and confidentiality protection~\cite{opcua_spec_security_2017}.
For client authentication, OPC UA allows anonymous access, a combination of username and password, a certificate, or an authentication token.
Furthermore, OPC UA servers can enforce access control for each node.
Correct configuration of authentication is indispensable, e.g., restricting anonymous access to non-critical servers~\cite{opcua_security_2018}.
Besides, default credentials of manufacturers must be changed.

To secure communication, OPC UA provides different message security modes (no security, integrity only, or integrity and confidentiality) as well as security policies predefining cryptographic algorithms for encryption and signatures~\cite{opcua_spec_security_2017}.
Depending on the message security mode, the client selects a security policy during the handshake to protect exchanged messages.
Notably, out of the seven available security policies, one provides no security, and two have been deprecated because of now insecure underlying cryptographic primitives.

Overall, while OPC UA provides strong security features~\cite{bsi_opcua_2017}, correct configuration of these security mechanisms is essential.
Otherwise, attackers might be able to access confidential data or compromise the OPC UA server.

\section{Assessing Security of Industrial Deployments}

The need to provide support for assessing the security of industrial deployments is emphasized by efforts for other industrial protocols.
Most prominently, different respective modules for the Metasploit Framework are available~\cite{caselli_security_2014}, ranging from modules for specific PLCs, e.g., Schneider Modicon or Siemens S7, over SCADA software, e.g., Sielco Sistemi Winlog or Measuresoft ScadaPro, to industrial protocols, e.g., Modbus, Profinet, or IEC 60870-5-104.
Furthermore, Masood et al.\ \cite{masood_swam_2011} recreated Stuxnet's attack vectors in Metasploit. 

Consequently, Metasploit is widely used for security assessments in industrial settings, e.g., in the production line of a pharmaceutical company~\cite{holik_pentest_2014}, to evaluate perimeter security effectiveness in supervisory and process control zones~\cite{combs_assessment_2016}, or to test industrial firewalls of a natural gas compressor~\cite{nguyen_strategy_2019}.
Likewise, NIST used Metasploit to assess the performance of industrial systems with security measures in place~\cite{candell_testbed_2015}.

\section{OPC UA Security Assessment}

\begin{figure}[t]
	\centerline{\includegraphics[trim=0 0.25cm 0 0, clip]{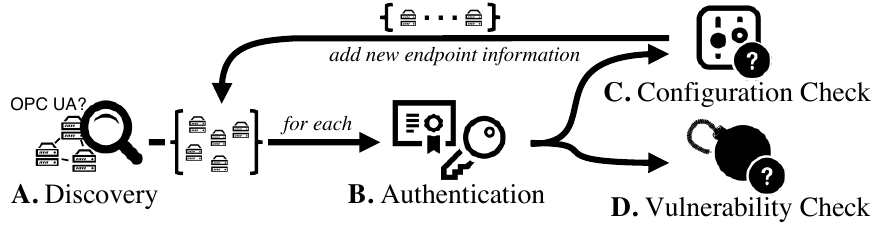}}
	\vspace{-0.3cm}
	\caption{Our approach for a network-based security assessment of OPC UA deployments (A) discovers OPC UA servers, (B) tests authentication, (C) derives server configuration, and (D) checks for vulnerabilities.}
	\vspace{-0.5cm}
	\label{fig}
\end{figure}

While OPC UA promises a high security level, actually achieving a secure OPC UA deployment depends on correct manual configuration (cf.\ Section~\ref{sec:background}).
Ensuring this for (own) OPC UA deployments requires network-based security assessments.
We use and extend the popular Metasploit Framework to aid in performing such tasks.
As illustrated in Figure~\ref{fig}, such an assessment requires several sequential steps.

\subsection{Discovery of OPC UA Servers}

As the first step of a network-based security assessment, we need to discover OPC UA servers in a local network.
By default, OPC UA uses TCP port 4840 for its binary protocol as well as TCP ports 80 (HTTP) and 443 (HTTPS) when acting as SOAP web service.
We extended the network scan functionality of Metasploit by an OPC UA handshake to verify that the discovered devices run OPC UA on said ports, thus obtaining a list of servers for subsequent assessment steps.

\subsection{Testing OPC UA Authentication \& Login Credentials}

Once an OPC UA server has been discovered, the next step is to test whether authentication is configured securely.
OPC UA allows for different modes of authentication (cf.\ Section~\ref{sec:background}).
Out of these modes, especially anonymous access and the widely-used username/password authentication are prone to configuration errors.
First, anonymous access (empty login credentials) might give away sensitive information on the server's configuration (see below) and imposes the risk of misconfigured access rights, leading to information leakage or even the risk of unauthorized write operations.
Second, username/password-based logins come with the risk of default (documented in manuals) or weak (easily guessable) login credentials.
To test for such problems, we added functionality to Metasploit to use the built-in \texttt{login\_scanner} for OPC UA deployments.
Furthermore, we created a list of OPC UA specific default credentials from openly accessible setup instructions and manuals. 
Following a similar route, we provide functionality to check whether the server accepts a self-signed client certificate without further validation.

\subsection{Deriving OPC UA Server Information \& Configuration}

If login to an OPC UA server is possible (either anonymous, using username/password, or certificate-based), we can connect to said server to obtain security-related information using our extension of Metasploit. 
First, information on available and known OPC UA servers can be used to broaden the scope of the assessment (cf.\ Figure~\ref{fig}).
Second, information such as \texttt{ApplicationUri} and \texttt{ProductUri} provide information on the used software and its version, easing the verification of patch policies and the check for potential vulnerabilities.
Third, different security-related configurations can be assessed, e.g., the use of an appropriate \texttt{SecurityLevel}, \texttt{MessageSecurityMode} enforcing encryption, security policy specified by the \texttt{PolicyUri}, and \texttt{TokenType} for authentication (cf.\ Section~\ref{sec:background}).
Finally, iterating through the server namespace of readable and writable nodes allows to identify misconfigured permissions (especially for anonymous authentication).
Combined, the information queryable after (anonymous) login to an OPC UA server allows to assess security of the underlying server configuration.

\subsection{Checking for Potential Vulnerabilities}

Besides configuration-based security assessment of an OPC UA deployment, we could actively test for susceptibility to known CVEs and other potential vulnerabilities.
To this end, Metasploit allows to run exploits for specific CVEs against a server.
Furthermore, some OPC UA servers (especially with anonymous access) are prone to a denial-of-service, i.e., attackers can exhaust the number of allowed active sessions.

\section{Outlook \& Conclusion}

To assist operators of OPC UA deployments in performing security assessments, we presented a network-based approach specific to OPC UA and extended the Metasploit Framework accordingly.
We verified our approach using both local test installations of different OPC UA implementations as well as public test servers.
Currently, we are further extending our toolset, especially by integrating more vulnerability checks.
In the future, we plan to validate our methods within industrial deployments and also cover other communication (e.g., publish/subscribe) and deployment (e.g., global discovery and proxy servers) paradigms.
With our work, we empower operators of OPC UA deployments to verify the security configuration of their systems and detect potential weaknesses.

\begin{footnotesize}
\noindent\\\textsc{Acknowledgments:}
This work has partly been funded by the Deutsche Forschungsgemeinschaft (DFG, German Research Foundation) under Germany's Excellence Strategy -- EXC-2023 Internet of Production -- 390621612.\par
\end{footnotesize}


\begin{thebibliography}{10}
\providecommand{\url}[1]{#1}
\csname url@samestyle\endcsname
\providecommand{\newblock}{\relax}
\providecommand{\bibinfo}[2]{#2}
\providecommand{\BIBentrySTDinterwordspacing}{\spaceskip=0pt\relax}
\providecommand{\BIBentryALTinterwordstretchfactor}{4}
\providecommand{\BIBentryALTinterwordspacing}{\spaceskip=\fontdimen2\font plus
\BIBentryALTinterwordstretchfactor\fontdimen3\font minus
  \fontdimen4\font\relax}
\providecommand{\BIBforeignlanguage}[2]{{%
\expandafter\ifx\csname l@#1\endcsname\relax
\typeout{** WARNING: IEEEtran.bst: No hyphenation pattern has been}%
\typeout{** loaded for the language `#1'. Using the pattern for}%
\typeout{** the default language instead.}%
\else
\language=\csname l@#1\endcsname
\fi
#2}}
\providecommand{\BIBdecl}{\relax}
\BIBdecl

\bibitem{pennekamp_towards_2019}
J.~Pennekamp \emph{et~al.}, ``{Towards an Infrastructure Enabling the Internet
  of Production},'' in \emph{IEEE ICPS}, 2019.

\bibitem{hemsley_history_2018}
K.~E. Hemsley and R.~E. Fisher, ``{History of Industrial Control System Cyber
  Incidents},'' Tech. Rep. INL/CON-18-44411-Revision-2, 2018.

\bibitem{opcua_specification_2017}
{OPC Foundation}, ``{OPC Unified Architecture Specification -- Part 1: Overview
  and Concepts},'' Version 1.04, 2017.

\bibitem{bsi_opcua_2017}
{Federal Office for Inform.\ Security}, ``{OPC UA Security Analysis},'' 2017.

\bibitem{opcua_security_2018}
{OPC Foundation}, ``{Practical Security Recommendations for building OPC UA
  Applications},'' Whitepaper, Version 3, 2018.

\bibitem{angeli_secure_2018}
C.~Angeli \emph{et~al.}, ``{Secure implementation of OPC UA for operators,
  integrators and manufacturers},'' Plattform Industrie 4.0, 2018.

\bibitem{fachkha_probing_2017}
C.~Fachkha \emph{et~al.}, ``{Internet-scale Probing of CPS: Inference,
  Characterization and Orchestration Analysis},'' in \emph{NDSS}, 2017.

\bibitem{opcua_spec_security_2017}
{OPC Foundation}, ``{OPC Unified Architecture Specification -- Part 2: Security
  Model},'' Version 1.04, 2017.

\bibitem{caselli_security_2014}
M.~Caselli and F.~Kargl, ``{A Security Assessment Methodology for Critical
  Infrastructures},'' in \emph{CRITIS}, 2014.

\bibitem{masood_swam_2011}
R.~Masood \emph{et~al.}, ``{Stuxnet Worm Analysis in Metasploit},'' in
  \emph{FIT}, 2011.

\bibitem{holik_pentest_2014}
F.~Holik \emph{et~al.}, ``{Effective penetration testing with Metasploit
  framework and methodologies},'' in \emph{IEEE CINTI}, 2014.

\bibitem{combs_assessment_2016}
M.~Combs-Ford, ``{Security Assessment of Industrial Control Supervisory and
  Process Control Zones},'' in \emph{SIGITE}, 2016.

\bibitem{nguyen_strategy_2019}
T.~D. Nguyen \emph{et~al.}, ``{A Strategy for Security Testing Industrial
  Firewalls},'' in \emph{ICSS}, 2019.

\bibitem{candell_testbed_2015}
R.~Candell \emph{et~al.}, ``{An Industrial Control System Cybersecurity
  Performance Testbed},'' NISTIR 8089, 2015.

\end{thebibliography}
\end{document}